# Interdependent Defense Games: Modeling Interdependent Security under Deliberate Attacks

**Hau Chan**
Department of Computer Science
Stony Brook University
hauchan@cs.stonybrook.edu

**Michael Ceyko**
Department of Computer Science
Harvard University
ceyko@fas.harvard.edu

**Luis E. Ortiz**
Department of Computer Science
Stony Brook University
leortiz@cs.stonybrook.edu

## Abstract

We propose *interdependent defense (IDD) games*, a computational game-theoretic framework to study aspects of the interdependence of risk and security in multi-agent systems under deliberate external attacks. Our model builds upon *interdependent security (IDS) games*, a model due to Heal and Kunreuther that considers the source of the risk to be the result of *a fixed randomized-strategy*. We adapt IDS games to model *the attacker's deliberate behavior*. We define the attacker's pure-strategy space and utility function and derive appropriate cost functions for the defenders. We provide a complete characterization of mixed-strategy Nash equilibria (MSNE), and design a simple *polynomial-time* algorithm for computing *all* of them, for an important subclass of IDD games. In addition, we propose a random-instance generator of (general) IDD games based on a version of the real-world Internet-derived Autonomous Systems (AS) graph (with around $27K$ nodes and $100K$ edges), and present promising empirical results using a simple learning heuristics to compute (approximate) MSNE in such games.

## 1 INTRODUCTION

Attacks carried out by hackers and terrorists over the last few years have led to increased efforts by both government and the private sector to create and adopt mechanisms to prevent future attacks. This effort has yielded a more focused research attention to models, computational and otherwise, that facilitate and help to improve (homeland) security for both physical infrastructure and cyberspace. In particular, there has been quite a bit of recent research activity in the general area of game-theoretic models for terrorism settings (see, e.g., Bier and Azaiez [2009] and Cárceles-Poveda and Tauman [2011]).

*Interdependent security (IDS) games* are one of the earliest models resulting from a game-theoretic approach to model security in non-cooperative environments composed of free-will self-interested individual decision-makers. Originally introduced and studied by economists Kunreuther and Heal [2003], IDS games model general abstract security problems in which an individual within a population considers whether to voluntarily invest in some protection mechanisms or security against a risk they may face, knowing that the cost-effectiveness of the decision depends on the investment decisions of others in the population because of transfer risks (i.e., the "bad event" may be transferable from a compromised individual to another).

In their work, Kunreuther and Heal [2003] provided several examples based on their economics, finance and risk management expertise. (We refer the reader to their paper for more detailed descriptions.) As a canonical example of the real-world relevance of IDS settings and the applicability of IDS games, Heal and Kunreuther [2005] used this model to describe problems such as airline baggage security. In their setting, individual airlines may choose to invest in additional complementary equipment to screen passengers' bags and check for hazards such as bombs that could cause damage to their passengers, planes, buildings, or even reputations. However, mainly due to the large amount of traffic volume, it is impractical for an airline to go beyond applying security checks to bags incoming from passengers and include checks to baggage or cargo transferred from other airlines. On the other hand, if an airline invests in security, they can still experience a bad event if the bag was transferred from an airline that does not screen incoming bags, rendering their investment useless.[1] Thus, we can see how

---

[1] Note that even if full screening were performed, the Christmas Day 2009 episode in Detroit [O'Connor and

the cost-effectiveness of an investment can be highly dependent on others' investment decisions. Another recent application of the IDS model is on container shipping transportation [Gkonis and Psaraftis, 2010]. They use the IDS model to study the effect of investment decision on container screening of ports have on their neighboring ports.

In this work, we build on the literature in IDS games. In particular, we adapt the model to situations in which the abstract "bad event" results from the *deliberate* action of an attacker. The "internal agents" (e.g., airlines and computer network users), whom we also often refer to as "defenders" or "sites," have the voluntary choice to individually invest in security to defend themselves against a direct or indirect offensive attack, modulo, of course, the cost-effectiveness to do so. A side benefit of explicitly modeling the attacker, as we do in our model, is that the probability of an attack results directly from the equilibrium analysis. Building IDS games can be hard because it requires *a priori* knowledge of the likelihood of an attack. Attacks of this kind are considered rare events and thus notoriously difficult to statistically estimate in general.

**Related Work.** Johnson et al. [2010] and Fultz and Grossklags [2009] independently developed non-cooperative game models similar to ours. Johnson et al. [2010] extend IDS games by modeling uncertainty about the source of the risk (i.e., the attacker) using a Bayesian game over risk parameters. Fultz and Grossklags [2009] propose and study a non-graphical game-theoretic model for the interactions between attackers and nodes in a network. In their model, each node in the network can decide on whether to contribute (by investment) to the overall safety of the network and/or to individual safety. The attackers can attack any number of nodes, but with each attack there is an increased probability that the attacker might get caught and suffer penalties or fines. Hence, while their game has IDS characteristics, it is technically not within the standard IDS game framework introduced by Heal and Kunreuther.

Most of the previous related work explore the realm of information security and are application/network specific (see Roy et al. [2010] for a survey on game theory application to network security). Past literature has largely focused on two-person (an attacker and a defender) games where the nodes in the network are regarded as a single entity (or a central defender). For example, Lye and Wing [2002] look at the interactions between an attacker and the (system) administrator using a two-player stochastic game. Recent work uses a Stackelberg game model in which the defender (or leader) commits to a mixed strategy to allocate resources to defend a set of nodes in the network, and the follower (or attacker) optimally allocates resources to attack a set of "targets" in the network given the leader's commitment [Jain et al., 2011, Kiekintveld et al., 2009, Korzhyk et al., 2010, 2011a,b].

Other recent work strive to understand the motivation of the attackers. For example, Liu [2003] focus on understanding the attacker's intent, objectives, and strategies and derive a (two-player) game-theoretical model based on these, while Cremonini and Nizovtsev [2006] use cost-benefit analysis (of attackers) to address the issue of the optimal amount of security (of the nodes in the network).

**Our Contribution.** We adapt the standard non-cooperative framework of IDS games to settings in which the source of the risk is the result of a deliberate, strategic decision by an external attacker. In particular, we design and propose *interdependent defense (IDD) games*, a new class of games that, in contrast to standard IDS games, model the attacker *explicitly*, while maintaining a core component of IDS systems: the potential *transferability* of the risk resulting from an attack. We note that the explicit modeling of risk transfer is an aspect of our model that has not been a focus of previous game-theoretic attacker-defender models of security discussed earlier.

We formally define and study IDD games in depth in Section 3. We present several results that *fully* characterize their NE. [2] We also provide a *polynomial-time* algorithm to compute *all* MSNE for the important subclass of IDD games in which there is only one attack, the defender nodes are *fully transfer-vulnerable* (i.e., investing in security does nothing to reduce their external/transfer risk) and transfers are *one-hop*. [3]

We note that considering a single attacker is a typical assumption in security settings (see previous work discussed earlier). It is also reasonable: We can view many attackers as a single attacker. Allowing at most one attack prevents immediate representational and computational intractability problems because the number of the attacker's (pure) strategies grows exponentially with the number of attacks. Finally, because the attacker has no fixed target, it is ineffective for the attacker to consider or go beyond plans of attacks involving multiple ($> 2$) transfers: such plans are complex, time consuming and costly.

Our computational results are significant and surprising because computing all NE in general IDS games is

---

Schmitt, 2009] serves as a reminder that transfer risk still exists.

[2] Due to space limitation, we omit proofs of our main technical results and instead refer the reader to the supplementary document for details [Chan et al., 2012].

[3] We note that the original IDS games were also *fully transfer-vulnerable* and assume one-hop transfers.

hard (i.e., the Nash-extension problem in general IDS games is NP-complete [Kearns and Ortiz, 2003]). [4] We do not know of any other *non-trivial* game for which there exists a *polynomial-time* algorithm to compute *all* NE except ours and the algorithm for uniform-transfer IDS games of Kearns and Ortiz [2003].

In Section 4, we provide experimental results from applying learning-in-games heuristics to compute approximate NE to both fixed and randomly-generated instances of IDD games, with at most one simultaneous attack and one-hop transfers, on a very large Internet AS graph ($\approx 27K$ nodes and $\approx 100K$ edges).

## 2 IDS GAMES

Each player $i$ in an IDS game has a choice to invest ($a_i = 1$) or not invest ($a_i = 0$). For each player $i$, $C_i$ and $L_i$ are the cost of investment and loss induce by the bad event, respectively. We define the ratio of the two parameters, the player's "cost-to-loss" ratio, as $\rho_i \equiv C_i/L_i$. Bad events can occur through both *direct* and *indirect* means. *Direct risk*, or *internal risk*, $p_i$ is the probability that player $i$ will experience a bad event because of direct contamination. The standard IDS model assumes that investing will completely protect the player from direct contamination; hence, internal risk is only possible when $a_i = 0$. *Indirect risk* $q_{ji}$ is the probability that player $j$ is directly "contaminated," does not experience the bad event but transfers it to player $i$ who ends up experiencing the bad event. There is an implicit global constraint on these parameters, by the axioms of probability: $p_i + \sum_{j=1}^n q_{ij} \leq 1$ for all $i$.

We now formally define a (directed) graphical games [Kearns, 2007, Kearns et al., 2001] version of IDS games, as first introduced by Kearns and Ortiz [2003]. Denote by $[n] \equiv \{1, \ldots, n\}$ the set of $n$ players. Note that the parameters $q_{ij}$'s induce a directed graph $G = ([n], E)$ such that $E \equiv \{(i,j) | q_{ij} > 0\}$. Let $\text{Pa}(i) \equiv \{j \mid q_{ji} > 0\}$ be the set of players that are *parents* of player $i$ in $G$ (i.e., the set of players that player $i$ is exposed to via transfers), and by $\text{PF}(i) \equiv \text{Pa}(i) \cup \{i\}$ the *parent family* of player $i$, which includes $i$. Denote by $k_i \equiv |\text{PF}(i)|$ the size of the parent family of player $i$. Similarly, let $\text{Ch}(i) \equiv \{j \mid q_{ij} > 0\}$ be the set of players that are *children* of player $i$ (i.e., the set of players to whom player $i$ can present a risk via transfer) and

$\text{CF}(i) \equiv \text{Ch}(i) \cup \{i\}$ the *(children) family* of player $i$, which includes $i$. The *probability that player $i$ is safe from player $j$*, as a function of player $j$'s decision, is

$$e_{ij}(a_j) \equiv a_j + (1 - a_j)(1 - q_{ji}) = (1 - q_{ji})^{1-a_j},$$

because if $j$ invests, then it is impossible for $j$ to transfer the bad event, while if $j$ does not invest, then $j$ either experiences the bad event or transfers it to another player, but never both.

Denote by $\mathbf{a} \equiv (a_1, \ldots, a_n) \in \{0,1\}^n$ the *joint action* of all $n$ players. Also denote by $\mathbf{a}_{-i}$ the joint-action of all players except $i$ and for any subset $I \subset [n]$ of players, denote by $\mathbf{a}_I$ the sub-component of the joint action corresponding to those players in $I$ only. We define $i$'s *overall safety* from *all* other players as $s_i(\mathbf{a}_{\text{Pa}(i)}) \equiv \prod_{j \in \text{Pa}(i)} e_{ij}(a_j)$ and equivalently his *overall risk* from *some* other players is $r_i(\mathbf{a}_{\text{Pa}(i)}) \equiv 1 - s_i(\mathbf{a}_{\text{Pa}(i)})$. Note that each players' external safety (and risk) is a direct function of its parents only, *not* all other players. From these definitions, we obtain player $i$'s overall cost, the cost of joint action $\mathbf{a} \in \{0,1\}^n$, corresponding to the (binary) investment decision of all players, is

$$M_i(a_i, \mathbf{a}_{\text{Pa}(i)}) \equiv a_i[C_i + r_i(\mathbf{a}_{\text{Pa}(i)})L_i] + (1 - a_i)[p_i + (1 - p_i)r_i(\mathbf{a}_{\text{Pa}(i)})]L_i .$$

## 3 IDD GAMES

In the standard IDS game model, investment in security does not reduce transfer risks. However, in some IDS settings (e.g., vaccination and cyber-security), it is reasonable to expect that security investments would include mechanisms to reduce transfer risks. This motivates our first modification to the traditional IDS games: allowing the investment in protection to not only makes us safe from direct attack but also partially reduce (or even eliminate) the transfer risk. We incorporate this factor by introducing a new real-valued parameter $\alpha_i \in [0, 1]$ representing the probability that a transfer of a potentially bad event will go *unblocked* by $i$'s security, assuming $i$ has invested. Thus, we redefine player $i$'s overall cost as [5]

$$M_i(a_i, \mathbf{a}_{\text{Pa}(i)}) \equiv a_i[C_i + \alpha_i r_i(\mathbf{a}_{\text{Pa}(i)})L_i] + (1 - a_i)[p_i + (1 - p_i)r_i(\mathbf{a}_{\text{Pa}(i)})]L_i .$$

Next, we introduce an additional player, the *attacker*, who *deliberately* initiates bad events. (So that now bad events are no longer "chance occurrences" without any strategic deliberation.) The attacker has a *target decision* for each player - a choice of attack ($b_i = 1$) or not attack ($b_i = 0$) player $i$. Hence, the attacker's pure strategy is denoted by the vector $\mathbf{b} \in \{0,1\}^n$. The game parameter $p_i$ implicitly "encodes" $b_i$ because $b_i = 0$ implies $p_i = 0$. Thus, we redefine $p_i \equiv p_i(b_i) \equiv b_i \widehat{p}_i$ so that player $i$ has *intrinsic risk* $\widehat{p}_i$, and only has *internal risk* if targeted (i.e, $b_i = 1$). The new parameter $\widehat{p}_i$ represents the (*conditional*) probability that an attack is successful at player $i$ *given* that player $i$ was directly targeted and did not invest in protection. The game parameter $q_{ij}$ "encodes" $b_i = 1$, because a prerequisite is that $i$ is targeted before it can transfer the bad event to $j$. We redefine $q_{ij} \equiv q_{ij}(b_i) \equiv b_i \widehat{q}_{ij}$ so that $\widehat{q}_{ij}$ is the intrinsic transfer probability from player $i$ to player $j$, independent of $b_i$. The new parameter $\widehat{q}_{ij}$ represents the (*conditional*) probability that an attack is successful at player $j$ *given* that it originated at player $i$, did not occur at $i$ but was transferred undetected to $j$. Note that just as it was the case with traditional IDS games, there is an implicit constraint on the risk-related parameters: $\widehat{p}_i + \sum_{j \in \text{Ch}(i)} \widehat{q}_{ij} \leq 1$, for all $i$. Because the $p_i$'s and $q_{ij}$'s depend on the attacker's action $\mathbf{b}$, so does the safety and risk functions. In particular, we now have

$$e_{ij}(a_j, b_j) \equiv a_j + (1-a_j)(1-b_j\widehat{q}_{ji}) = (1-\widehat{q}_{ji})^{b_j(1-a_j)},$$

---

[4] To put our computational contributions in context, note that deciding whether a game has a PSNE is in general NP-complete (see, e.g., Gilboa and Zemel [1989], Gottlob et al. [2005]), and computing an MSNE is PPAD-complete, even in two-player games (see, e.g., Chen et al. [2009] and Daskalakis et al. [2009]). Also, computing *all* MSNE is rarely achieved and counting-related problems are often #P-complete (see, e.g., [Conitzer and Sandholm, 2008]).

[5] A similar extension was also proposed independently by Heal and Kunreuther [2007].

$s_i(\mathbf{a}_{\text{Pa}(i)}, \mathbf{b}_{\text{Pa}(i)}) \equiv \prod_{j \in \text{Pa}(i)} e_{ij}(a_j, b_j) \equiv 1 - r_i(\mathbf{a}_{\text{Pa}(i)}, \mathbf{b}_{\text{Pa}(i)})$. Hence, for each player $i$, the *cost* function becomes

$$M_i(a_i, \mathbf{a}_{\text{Pa}(i)}, b_i, \mathbf{b}_{\text{Pa}(i)}) \equiv a_i[C_i + \alpha_i r_i(\mathbf{a}_{\text{Pa}(i)}, \mathbf{b}_{\text{Pa}(i)})L_i] \\ + (1-a_i)[b_i\widehat{p}_i + (1-b_i\widehat{p}_i)r_i(\mathbf{a}_{\text{Pa}(i)}, \mathbf{b}_{\text{Pa}(i)})]L_i.$$

We assume the attacker wants to cause as much damage as possible. One possible *utility/payoff* function $U$ quantifying the objective of the attacker is

$$U(\mathbf{a}, \mathbf{b}) \equiv \sum_{i=1}^n M_i(\mathbf{a}_{\text{PF}(i)}, \mathbf{b}_{\text{PF}(i)}) - a_iC_i - b_iC_i^0.$$

which adds the expected players costs (for targeted and transferred bad events) over all players, minus $C_i^0$, the attacker's own "cost" to target player $i$.

### 3.1 MIXED STRATEGIES IN IDD GAMES

For all player $i$, denote by $x_i$ the *mixed strategy of player i*: the probability that player $i$ invests. Similarly, $y$ denotes the joint probability mass function (PMF) corresponding to the *attacker's mixed strategy* so that for all $\mathbf{b} \in \{0,1\}^n$, $y(\mathbf{b})$ is the probability that the attacker executes joint-attack vector $\mathbf{b}$.

Denote the marginal PMF over a subset $I \subset [n]$ of the internal players by $y_I$ such that for all $\mathbf{b}_I$, $y_I(\mathbf{b}_I) \equiv \sum_{\mathbf{b}_{-I}} y(\mathbf{b}_I, \mathbf{b}_{-I})$ is the (marginal) probability that the attacker chooses a joint-attack vector in which the sub-component decisions corresponding to players in $I$ are as in $\mathbf{b}_I$. Denote simply by $y_i \equiv y_{\{i\}}(1)$ the marginal probability that the attacker chooses an attack vector in which player $i$ is directly targeted. Slightly abusing notation, we redefine the function $e_{ij}$ (i.e., how safe $i$ is from $j$), $s_i$ and $r_i$ (i.e., the overall transfer safety and risk, respectively) as $e_{ij}(x_j, b_j) \equiv x_j + (1-x_j)(1-b_j\widehat{q}_{ji})$, $s_i(\mathbf{x}_{\text{Pa}(i)}, \mathbf{b}_{\text{Pa}(i)}) \equiv \prod_{j \in \text{Pa}(i)} e_{ij}(x_j, b_j)$,

$$s_i(\mathbf{x}_{\text{Pa}(i)}, y_{\text{Pa}(i)}) \equiv \sum_{\mathbf{b}_{\text{Pa}(i)}} y_{\text{Pa}(i)}(\mathbf{b}_{\text{Pa}(i)}) s_i(\mathbf{x}_{\text{Pa}(i)}, \mathbf{b}_{\text{Pa}(i)}),$$

and $r_i(\mathbf{x}_{\text{Pa}(i)}, y_{\text{Pa}(i)}) \equiv 1 - s_i(\mathbf{x}_{\text{Pa}(i)}, y_{\text{Pa}(i)})$.

In general, the *expected* cost of protection to site $i$, with respect to a joint mixed-strategy $(\mathbf{x}, y)$, can be expressed as

$$M_i(x_i, \mathbf{x}_{\text{Pa}(i)}, y_{\text{PF}(i)}) \equiv x_i[C_i + \alpha_i r_i(\mathbf{x}_{\text{Pa}(i)}, y_{\text{Pa}(i)})L_i] + \\ (1-x_i)[\widehat{p}_i f_i(\mathbf{x}_{\text{Pa}(i)}, y_{\text{PF}(i)}) + r_i(\mathbf{x}_{\text{Pa}(i)}, y_{\text{Pa}(i)})]L_i ,$$

where $f_i(\mathbf{x}_{\text{Pa}(i)}, y_{\text{PF}(i)}) \equiv$

$$\sum_{\mathbf{b}_{\text{PF}(i)}} y_{\text{PF}(i)}(\mathbf{b}_{\text{PF}(i)}) \, b_i s_i(\mathbf{x}_{\text{Pa}(i)}, \mathbf{b}_{\text{Pa}(i)}) .$$

The expected payoff of the attacker is

$$U(\mathbf{x}, y) \equiv \sum_{i=1}^n M_i(\mathbf{x}_{\text{PF}(i)}, y_{\text{PF}(i)}) - x_iC_i - y_iC_i^0.$$

Let $\widehat{\Delta}_i \equiv \rho_i/\widehat{p}_i \equiv \frac{C_i}{L_i\widehat{p}_i}$ and $\widehat{s}_i(\mathbf{x}_{\text{Pa}(i)}, y_{\text{PF}(i)}) \equiv f_i(\mathbf{x}_{\text{Pa}(i)}, y_{\text{PF}(i)}) + \frac{1-\alpha_i}{\widehat{p}_i} r_i(x_{\text{Pa}(i)}, y_{\text{Pa}(i)})$. The best-response correspondence of defender $i$ is then

$$\mathcal{BR}_i(\mathbf{x}_{\text{Pa}(i)}, y_{\text{PF}(i)}) \equiv \begin{cases} \{1\}, & \text{if } \widehat{s}_i(\mathbf{x}_{\text{Pa}(i)}, y_{\text{PF}(i)}) > \widehat{\Delta}_i, \\ \{0\}, & \text{if } \widehat{s}_i(\mathbf{x}_{\text{Pa}(i)}, y_{\text{PF}(i)}) < \widehat{\Delta}_i, \\ [0,1], & \text{if } \widehat{s}_i(\mathbf{x}_{\text{Pa}(i)}, y_{\text{PF}(i)}) = \widehat{\Delta}_i. \end{cases}$$

The best-response correspondence for the attacker is simply $\mathcal{BR}_0(\mathbf{x}) \equiv \arg\max_y U(\mathbf{x}, y)$.

**Definition 1** *A joint mixed-strategy* $(\mathbf{x}^*, y^*)$ *is a mixed-strategy Nash equilibrium (MSNE) of an IDD game if (1) for all* $i \in [n]$, $x_i^* \in \mathcal{BR}_i(\mathbf{x}_{Pa(i)}^*, y_{\text{PF}(i)}^*)$ *and (2)* $y^* \in \mathcal{BR}_0(\mathbf{x}^*)$. *If* $(\mathbf{x}^*, y^*)$ *corresponds to a (deterministic) joint action then we call the MSNE a pure-strategy Nash equilibrium (PSNE).*

### 3.2 MODEL ASSUMPTIONS

Note that the attacker has in principle an *exponential* number of pure strategies! This affords the attacker unrealistic amount of power. Hence, we need restriction on the attacker's power. The simplest way is to allow at most a single simultaneous attack. We can weaken this assumption to allow the attacker at most $K$ simultaneous attacks. Even then, the number of pure strategies will grow *exponentially* in the number of potential attacks, which still renders the attacker's pure-strategy space unrealistic, especially on a very large network with twenty-thousand nodes. Worst-case, we need to consider up to $2^n$ number of pure strategies for $K$ attacks as $K$ goes to $n$.

**Assumption 1** *The set of pure strategies of the attacker is $\mathcal{B} = \{\mathbf{b} \in \{0,1\}^n \mid \sum_{i=1}^n b_i \leq 1\}$.*

The following assumptions are on the game *parameters*. The next assumption states that every site's investment cost is positive and (strictly) smaller than the *conditional* expected *direct* loss if the site were to be attacked directly ($b_i = 1$); that is, if a site knows that an attack is directed against it, the site will prefer to invest in security, unless the *external risk* is too high. This assumption is reasonable because otherwise the player will never invest regardless of what other players do (i.e., not investing would be a dominant strategy).

**Assumption 2** *For all sites $i \in [n]$, $0 < C_i < \widehat{p}_i L_i$.*

The next assumption states that, for all sites $i$, the attacker's cost to attack $i$ is positive and (strictly) smaller than the expected loss (i.e., gains from the perspective of the attacker) achieved if an attack initiated at site $i$ is successful, either directly at $i$ or at one of its children (after transfer); that is, if an attacker knows that an attack is rewarding (or able to obtain a positive utility), it will prefer to attack some nodes in the network. This assumption is reasonable; otherwise the attacker will never attack regardless of what other players do (i.e., not attacking would be a dominant strategy, leading to an easy problem to solve).

**Assumption 3** *For all sites $i \in [n]$, $0 < C_i^0 < \widehat{p}_i L_i + \sum_{j \in Ch(i)} \widehat{q}_{ij} \alpha_j L_j$.*

**PSNE of IDD Games.** It turns out that under these three assumptions, there is no PSNE in IDD games. This is typical of attacker-defender settings. The following proposition eliminates PSNE as a universal solution concept for natural IDD games in which at most one attack is possible. The main significance of this result is that it allows us to concentrate our efforts on the much harder problem of computing MSNE.

**Proposition 1** *No IDD game in which Assumptions 1, 2 and 3 hold has a PSNE.*

### 3.3 MSNE OF IDD GAMES

We first consider the IDD games under the assumption that no protection from transfer risk, which is used in the original IDS games. Note that, throughout this subsection, we will be using the assumptions we made earlier (in additional to the following assumption).

**Assumption 4** *For all internal players $i \in N$, the probability that player $i$'s investment in security does not protect the player from transfers, $\alpha_i$, is 1.*

**Definition 2** *We say an IDD game is* transfer-vulnerable *if Assumption 4 holds. We say an IDD game is a* single simultaneous attack *game if Assumption 1 holds (i.e., at most one attack is possible).*

Assumption 1, in the context of mixed strategies, implies the probability of no attack $y_0 \equiv 1 - \sum_i^n y_i$. Assumptions 1 and 4 greatly simplify the best-response condition of the internal players because now $\widehat{s}_i(\mathbf{x}_{\text{Pa}(i)}, y_{\text{PF}(i)}) = y_i$. Let $L_i^0(x_i) \equiv (1-x_i)(\widehat{p}_i L_i + \sum_{j \in \text{Ch}(i)} \widehat{q}_{ij} L_j)$. It will also be convenient to denote by $\overline{L}_i^0 \equiv L_i^0(0) = \widehat{p}_i L_i + \sum_{j \in \text{Ch}(i)} \widehat{q}_{ij} L_j$, so that we can express $L_i^0(x_i) = (1-x_i)\overline{L}_i^0$, to highlight that $L_i^0$ is a linear function of $x_i$. Similarly, it will also be convenient to let $M_i^0(x_i) \equiv L_i^0(x_i) - C_i^0$, and denote $\overline{M}_i^0 \equiv M_i^0(0) = \overline{L}_i^0 - C_i^0$. Let $\eta_i^0 \equiv C_i^0/\overline{L}_i^0$. The best-response condition of the attacker also simplifies under the same assumptions because now $U(\mathbf{x}, y) = \sum_{i=1}^n y_i M_i^0(x_i)$. Assumption 3 is reasonable in our new context because, under Assumption 4, if there were a player $i$ with $\eta_i^0 > 1$, the attacker would never attack $i$, and as a result player $i$ would never invest. In that case, we can safely remove $j$ from the game, without any loss of generality.

We now characterize the space of MSNE in IDD games, which will immediately lead to a polynomial-time algorithm for computing *all* MSNE.

**Characterization.** The characterization starts by partitioning the space of games into three, based on whether $\sum_{i=1}^n \widehat{\Delta}_i$ is (1) $<$, (2) $=$, or (3) $>$ than 1. The rationale behind this is that now the players are indifferent between investing or not investing when $y_i = \widehat{\Delta}_i$, by the best-response correspondence the attacker's mixed strategy is restricted. The following result fully characterizes the set of MSNE in single simultaneous attack transfer-vulnerable IDD games.

**Proposition 2** *The joint mixed-strategy $(\mathbf{x}^*, y^*)$ is an MSNE of a single simultaneous attack transfer-vulnerable IDD game in which*

1. $\sum_{i=1}^n \widehat{\Delta}_i < 1$ *if and only if (1) $1 > y_0^* = 1 - \sum_{i=1}^n \widehat{\Delta}_i > 0$, and (2) for all $i$, $y_i^* = \widehat{\Delta}_i > 0$ and $0 < x_i^* = 1 - \eta_i^0 < 1$.*

2. $\sum_{i=1}^n \widehat{\Delta}_i = 1$ *if and only if (1) $y_0^* = 0$, and (2) for all $i$, $y_i^* = \widehat{\Delta}_i > 0$ and $x_i^* = 1 - \frac{v + C_i^0}{\overline{L}_i^0}$ with $0 \leq v \leq \min_{i \in [n]} \overline{M}_i^0$.*

3. $\sum_{i=1}^n \widehat{\Delta}_i > 1$ *if and only if (1) $y_0^* = 0$, and (2) there exists a non-singleton, non-empty subset $I \subset [n]$, such that $\min_{i \in I} \overline{M}_i^0 \geq \max_{k \notin I} \overline{M}_k^0$ if $I \neq [n]$, and the following holds: (a) for all $k \notin I$, $x_k^* = 0$ and $y_k^* = 0$, (b) for all $i \in J \equiv \arg\min_{i \in I} \overline{M}_i^0$, $x_i^* = 0$ and $0 \leq y_i^* \leq \widehat{\Delta}_i$,* and in addition, $\sum_{i \in J} y_i^* = 1 - \sum_{t \in I-J} \widehat{\Delta}_i$; and (c) for all $i \in I - J$, $y_i^* = \widehat{\Delta}_i$ and $0 < x_i^* = 1 - \frac{\min_{t \in I} \overline{M}_t^0 + C_i^0}{\overline{L}_i^0} < 1$.

As proof sketch, we briefly state that the proposition follows from the restrictions imposed by the model parameters and their implication to indifference and monotonicity conditions. We also mention that the third case in the proposition implies that if the $\overline{M}_l^0$'s form a complete order, then the last condition stated in that case allows us to search for a MSNE by exploring only $n - 2$ sets, vs. $2^{n-2}$ if done naively. [6]

We now discuss properties of the characterization.

**Security investment characteristics of MSNE.** At equilibrium $\mathbf{x}^*$, if $x_i^* > 0$, the probability of *not* investing is proportional to $C_i^0$ and *inversely* proportional to $\widehat{p}_i \overline{L} + \sum_{j \in \text{Ch}(i)} \widehat{q}_{ij} L_j$. It is kind of reassuring the at equilibrium, which is the (almost-surely) unique stable outcome of the system, the probability of investing increases with the potential loss a player's non-investment decision could cause to the system. Hence, behavior in a stable system implicitly "forces" all players to indirectly account for or take care of their own children. This may sound a bit paradoxical at first given that we are working within "noncooperative" setting and each player's cost function is only dependent on the investment decision of the player's *parents*. Interestingly, the existence of the attacker in the system is inducing an (almost-surely) unique stable outcome in which an implicit form of "cooperation" occurs. A defenders's best response is independent of their parents, the source of transfer risk, if investment in security does nothing to protect that player from transfers (i.e., $\alpha_i = 1$). This makes sense because the player cannot control the transfer risk.

**Relation to network structure.** How does the network structure and the equilibrium relate? As seen above, the values of the equilibrium strategy of each player depend on information from the attacker, the player and the player's children. From the discussion in the last paragraph, a player's probability of investing at the equilibrium increases with the expected loss sustained from a "bad event" occurring as a result of a transfer from a player to the player's children.

Let us explore this last point further by considering the case of uniform-transfer probabilities (also studied by Kunreuther and Heal [2003] and Kearns and Ortiz [2003]). In that case, transfer probabilities are only a function of the source, not the destination: $\widehat{q}_{ij} \equiv \widehat{\delta}_i$.

The expression for the equilibrium probabilities of those players who have a positive probability of investing would simplify to $x_i^* = 1 - \frac{v + C_i^0}{\widehat{p}_i L_i + \delta_i \sum_{j \in \text{Ch}(i)} L_j}$, for some constant $v$. The last expression suggests that $\sum_{j \in \text{Ch}(i)} L_j$ differentiates the probability of investing between players. That would suggests the larger the number of children the larger the probability of investing. A scenario that seems to further lead us to that conclusion is when we make further assumptions (homogeneous system: first studied in the original IDS paper): $L_i \equiv L$, $\widehat{p}_i \equiv \widehat{p}$, $\delta_i \equiv \delta$, and $C_i^0 \equiv C^0$ [7] for all players. Then, we would get $x_i^* = 1 - \frac{v + C^0}{L(\widehat{p} + \delta |\text{Ch}(i)|)}$. So the probability of *not* investing is inversely proportional to the *number* of children the player has.

**On the attacker's equilibrium strategy.** The support of the attacker, $I^* \equiv \{i \mid y_i^* > 0\}$, at equilibrium has the following properties: (1) players for which the attacker's cost-to-expected-loss is higher are "selected" first in the algorithm; (2) if the size of that set is $t$, and there is a lower bound on $\widehat{\Delta}_i > \widehat{\Delta}$, and $\sum_{i=1}^n \widehat{\Delta}_i > 1$, then $t < 1/\widehat{\Delta}$ is an upper-bound on the number of players that could potentially be attacked; (3) if we have a game with homogeneous parameters, then the probability of an attack will be uniform over that set $I^*$; and (4) all but one of the players in that set $I^*$ invest in security with some non-zero probability (almost surely).

### 3.4 COMPUTING MSNE EFFICIENTLY

We now describe an algorithm to compute *all* MSNE in single simultaneous attack transfer-vulnerable IDD games that falls off the Proposition 2. We begin by noting that the equilibrium in the case of IDD games with $\sum_{i=1}^n \widehat{\Delta}_i \leq 1$, corresponding to cases 1 and 2 of the proposition, has essentially an analytic closed-form. Hence, we concentrate on the remaining and most realistic case in large-population games of $\sum_{i=1}^n \widehat{\Delta}_i > 1$. We start by sorting the indices of the internal players in descending order based on the $\overline{M}_i^0$'s. Let $\text{Val}(l)$ and $\text{Idx}(l)$ be the $l$th value and index in the resulting sorted list, respectively. Find $t$ such that $1 - \widehat{\Delta}_{\text{Idx}(t)} \leq \sum_{l=1}^{t-1} \widehat{\Delta}_{\text{Idx}(l)} < 1$. Let $k = \arg\max\{l \geq t \mid \text{Val}(l) = \text{Val}(t)\}$ (i.e., continue down the sorted list of values until a change occurs). For $i = 1, \ldots, t-1$, let $l = \text{Idx}(i)$ and set $x_l^* = 1 - \frac{\text{Val}(t) + C_l^0}{\overline{L}_l^0}$ and $y_l^* = \widehat{\Delta}_l$. For $i = k+1, \ldots, n$, let $l = \text{Idx}(i)$ and set $x_l^* = 0$ and $y_l^* = 0$. For $i = t, \ldots, k$, let $l = \text{Idx}(i)$ and set $x_l^* = 0$. Finally, represent the simplex defined by the following constraints: for $i = t, \ldots, k$, let $l = \text{Idx}(i)$ and $0 \leq y_l^* \leq \widehat{\Delta}_l$;

---

[6] In fact, it turns out a complete order is not actually necessary because we can safely move all defenders with the same value of $\overline{M}_i^0$ in a group as a whole inside or outside the set $I$ referred to in that case of the proposition.

[7] Note that this does not mean that the expected loss caused by a player that does not invest but is attacked, $L(\widehat{p} + \delta |\text{Ch}(i)|)$, is the same for all players.

$\sum_{i=t}^{k} y_{\text{Idx}(i)}^* = 1 - \sum_{i=1}^{t-1} \widehat{\Delta}_{\text{Idx}(i)}$. The running time of the algorithm is $O(n \log n)$ (because of sorting). As mentioned earlier, the theorem is significant because we do not know of any other non-trivial class of games that have algorithms to compute all Nash equilibria in polynomial time, except ours and the uniform IDS games [Kearns and Ortiz, 2003].

**Theorem 1** *There exists a polynomial-time algorithm to compute all MSNE of a single simultaneous attack transfer-vulnerable IDD game.*

Note that the case in which $\sum_{i=1}^{n} \widehat{\Delta}_i = 1$ has (Borel) measure zero and is quite brittle (i.e., adding or removing a player breaks the equality). For the case in which $\sum_{i=1}^{n} \widehat{\Delta}_i > 1$, if the value of the $\overline{M}_i^0$'s are distinct, [8] then there is a unique MSNE!

## 4 EXPERIMENTS

In the previous section, we established the theoretical characteristics and computational tractability of single simultaneous attack IDD games with the highest transfer vulnerability parameter: $\alpha_i = 1$. In this section, partly motivated by security problems in cyberspace, we concentrate instead on empirically evaluating the other extreme of transfer vulnerability: games with low $\alpha_i$ values (i.e., near 0), so that investing in security considerably reduces the transfer risk.

Our main objectives for the experiments presented here are (1) to demonstrate that a simple heuristic, *best-response-gradient dynamics (BRGD)*, is practically effective in computing an (approximate) MSNE in a very large class of IDD games with realistic Internet-scale network graphs in a reasonable amount of time for cases in which the transfer vulnerabilities $\alpha_i$'s are low; and (2) to explore the general structural and computational characteristics of (approximate) MSNE in such IDD games, including their dependence on the underlying network structure of the game (and approximation quality).

BRGD is a well-known technique from the work on learning in games [Fudenberg and Levine, 1999, Singh et al., 2000, Kearns and Ortiz, 2003, Heal and Kunreuther, 2005, Kearns, 2005]. Here, we use BRGD as a tool to *compute* an $\epsilon$-approximate MSNE, which is a joint mixed strategy with the property that the gain in utility (or reduction in cost) of any individual from unilaterally deviating from their prescribed mixed-strategy is no larger than $\epsilon$; a 0-approximate MSNE is an exact MSNE.

[8] Distinct $\overline{M}_i^0$'s for the set of defenders at which the sum goes over one is sufficient to guarantee unique MSNE.

Table 1: Internet Games' Model Parameters

| Model Parameters | Fixed: U = 0.5  Random: U ~ Uniform([0,1]) |
|---|---|
| $\alpha_i$ | $U/20$ |
| $L_i$ | $10^8 + (10^9) * U$ |
| $C_i$ | $10^5 + (10^6) * U$ |
| $\widehat{p}_i$ | $0.9 * \frac{\overline{p}_i}{\overline{p}_i + \sum_{k \in \text{Ch}(i)} \overline{q}_{ik}}$ |
| $\widehat{q}_{ij}$ | $0.9 * \frac{\overline{q}_{ij}}{\overline{p}_i + \sum_{k \in \text{Ch}(i)} \overline{q}_{ik}}$ |
| $z_i$ | $0.2 + U/5$ |
| $\overline{p}_i$ | $0.8 + U/10$ |
| $\widetilde{q}_{ij}$ | $z_i \frac{|\text{Ch}(j)| + |\text{Pa}(j)|}{\sum_{k \in \text{Ch}(i)} |\text{Ch}(k)| + |\text{Pa}(k)|}$ |
| $C_i^0$ | $10^6$ |

We obtained the latest version (March 2010 at the time) of the real structure and topology of the Autonomous Systems (AS) in the Internet from DIMES (netdimes.org) [Shavitt and Shir, 2005]. The AS-level network has $27,106$ nodes (683 isolated) and $100,402$ directed edges; the graph length (diameter) is $6,253$, the density (number of edges divided by number of possible edges) is $1.9920 \times 10^{-5}$, and the average (in and out) degree is $3.70$, with $\approx 76.93\%$ and $2.59\%$ of the nodes having zero indegree and outdegree, respectively. *All the IDD games in the experiments presented in this section have this network structure.*

For simplicity, we call *Internet games* the class of IDD games with the AS-level network graph and low $\alpha_i$ values. We considered various settings for model parameters of Internet games: a single instance with specific fixed values; and several instances generated at random (see Table 1 for details). The attacker's cost-to-attack parameter for each node $i$ is always held constant: $C_i^0 = 10^6$. For each run of each experiment, we ran BRGD with randomly-generated initial conditions (i.e., random initializations of the players' mixed strategies): $x_i \sim \text{Uniform}([0, 1])$, i.i.d. for all $i$, and $y$ is a probability distribution generated uniformly at random, and independent of $\mathbf{x}$, from the set of all probability mass functions over $n + 1$ events. [9] The initialization of the transfer-probability parameters of a node essentially gives higher transfer probability to children with high (total) degree (because they are potentially "more popular"). The initialization also enforces $\widehat{p}_i + \sum_{j \in \text{Ch}(i)} \widehat{q}_{ij} = 0.9$. Other initializations are possible but we did not explore then here.

### 4.1 COMPUTING $\epsilon$-MSNE USING BRGD

Given the lack of theoretical guarantees on the convergence rate of BRGD, we began our empirical study by evaluating the convergence and computation/running-time behavior of BRGD on Internet games. We ran ten simulations for each $\epsilon$ value and recorded the number of iterations until convergence (up to $2,000$ iterations). Figure 1 presents the number of iterations taken by BRGD to compute an $\epsilon$-MSNE as a function of $\epsilon$. All simulations in this experiment converged (except for $\epsilon = 0.001$: 2 and all of the runs for single and randomly-generated instances, respectively, did not). Each iteration took roughly 1-2 sec. (wall clock). Hence, we can use BRGD to consistently compute an $\epsilon$-MSNE of a 27K-players Internet game in a few seconds!

[9] Recall the probability of no attack $y_0 = 1 - \sum_{i=1}^{n} y_i$.

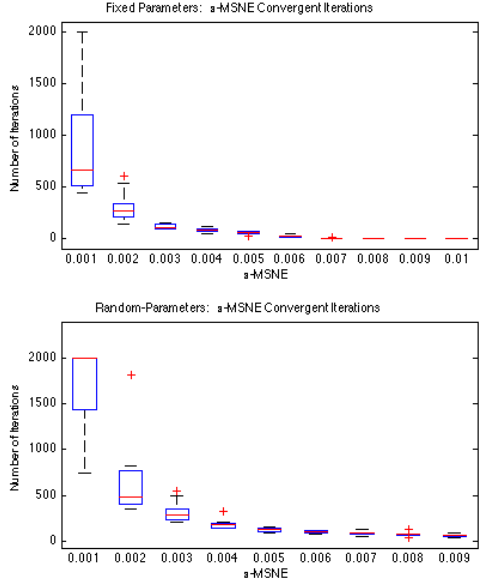

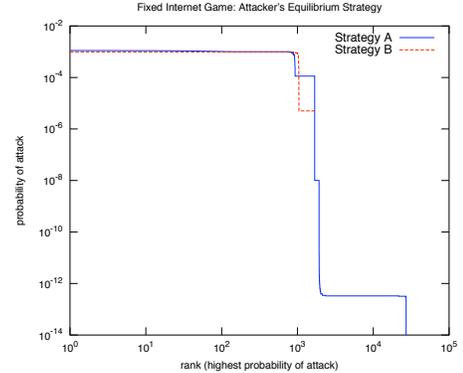

Figure 2: **Attacker's Equilibrium Strategy on an Internet Game Instance (Fixed):** The graph shows the values of $y_i^* > 0$ for each node $i$, sorted in decreasing order (in log-log scale), for attacker's Strategy A (blue/denser-dots line) and Strategy B (red/sparser-dots line) at an MSNE of the single instance of the Internet game.

Figure 1: **Convergence Rate of Learning Dynamics:** The plots above present the number of iterations of BRGD as a function of $\epsilon$ under the two experimental conditions: Internet games with fixed (top) and randomly-generated parameters (bottom). Applying MSE regression to the top and bottom graphs, we obtain a functional expression for the number of iterations $N^F(\epsilon) = 0.00003\epsilon^{-2.547}$ ( $R^2 = 0.90415$) and $N^R(\epsilon) = 0.0291\epsilon^{-1.589}$ ($R^2 = 0.9395$), respectively (i.e., low-degree polynomials of $1/\epsilon$).

### 4.2 CHARACTERISTICS OF THE $\epsilon$-MSNE

We now concentrate on the empirical study of the *structural* characteristics of the $\epsilon$-MSNE found by BRGD. We experimented on both the single and randomly-generated Internet game instances. We discuss the typical behavior of the attacker and the sites in an $\epsilon$-MSNE, and the typical relationship between $\epsilon$-MSNE and network structure.

#### 4.2.1 A Single Internet Game

We first studied the characteristics of the $\epsilon$-MSNE of a single Internet game instance. The only source of randomness in these experiments comes from BRGD's initial conditions (i.e., the initialization of the mixed strategies $\mathbf{x}$ and $\mathbf{y}$). BRGD consistently found *exact* MSNE (i.e., $\epsilon = 0$) in *all* runs.

**Players' equilibrium behavior.** In fact, we consistently found that the attacker always displays only two types of "extreme" equilibrium behavior, corresponding to the two kinds of MSNE BRGD found for the single Internet game: place positive probability of a direct attack to either *almost all* nodes (Strategy A) or a *small subset* (Strategy B). Figure 2 shows a plot of the typical probability of direct attack for those two equilibrium strategies for the attacker when BRGD stops. In both strategies, a relatively small number of nodes (about 1K out of 27K) have a reasonably *high* (and near *uniform*) probability of direct attack. In Strategy A, however, *every* node has a positive probability of being the target of a direct attack, albeit relatively very low for most; this is contrast to Strategy B where *most* nodes are fully immune from a direct attack. Interestingly, *none* of the nodes invest in either MSNE: $x_i^* = 0$ for all nodes $i$. Thus, in this particular Internet game instance, *all* site nodes are willing to risk an attack!

**Relation to network structure.** We found that the nodes with (relatively) high probability of direct attack are at the "fringe" of the graph (i.e., have low or no degree). In Strategy A, fringe nodes (with mostly 0 or 1 outdegree) have relatively higher probability of direct attack than nodes with higher outdegree. Similarly, in Strategy B, the small subset of nodes that are potential target of a direct attack have relatively low outdegree (mostly 0, and 0.0067 on average; this is in contrast to the average outdegree of 3.9639 for the nodes immune from direct attack). In short, we consistently found that the nodes with low outdegree

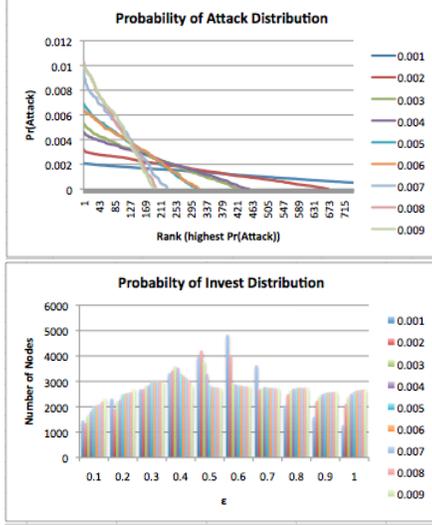

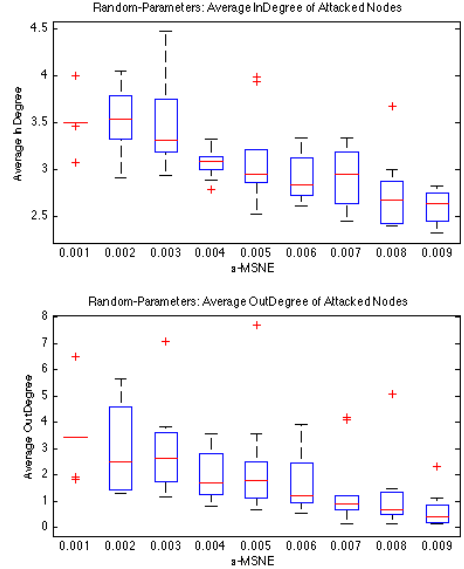

Figure 3: **Attacker's and Site's $\epsilon$-MSNE Strategies for Randomly-Generated Internet Games:** The graphs show the empirical distributions of the probability of attack (top) and histograms of the probability of investment (bottom), for different $\epsilon$-value conditions encoded in the right-hand side of the plots (i.e., from 0.001 to 0.009). In both graphs, the distributions and histograms found for each $\epsilon$ value considered are drawn in the same corresponding graph superimposed. The top graph plots the distribution of $y_i$ where the nodes are ordered decreasingly based on the $y_i$ value. The bottom bar graph shows histograms of the probability of investing in defense/security measures based on the following sequence of 10 ranges partitioning the unit interval: $([0, 0.1], (0.1, 0.2], ..., (0.9, 1])$.

are more likely to get attacked directly in the single Internet game instance.

### 4.2.2 Randomly-Generated Internet Games

We now present results from experiments on randomly-generated instances of the Internet game, a single instance for each $\epsilon \in \{0.001, 0.002, \ldots, 0.009\}$. For simplicity, we present the result of a single BRGD run on each instance.[10]

**Behavior of the players.** Figure 3 shows plots of the attacker's probability of direct attack and histograms of the nodes's probability of investment in a typical run of BRDG on each Internet game instance randomly-generated for each $\epsilon$ value. The plots suggest that approximate MSNE found by BRGD is quite sensitive to the $\epsilon$ value: as $\epsilon$ decreases, the attacker tends to "spread the risk" by selecting a larger set of nodes as potential targets for a direct attack, thus lowering the probability of a direct attack on any individual node; the nodes, on the other hand, tend to deviate from (almost) fully investing and (almost) not investing to a more uniform mixed strategy (i.e., investing or not investing with roughly equal probability). A possible reason for this is that as more nodes become potential targets of a direct attack, more nodes with initial mixed strategies close to the "extreme" (i.e., very high or very low probabilities of investing) will move closer to a more uniform (and thus less "predictable") investment distribution.

---
[10]While the results presented here are for a single instance of the Internet game for each $\epsilon$, the results are typical of multiple instances. Our observations are robust to the experimental randomness in both the Internet game parameters and the initialization of BRGD. For the sake of simplicity of presentation, we discuss results based on a single instance of the Internet game, and in some cases based on a single BRGD run. Note that, for each $\epsilon$ value we considered, the Internet game parameters remain constant within different BRGD runs. BRGD always converged within 2,000 iterations (except 6 runs for $\epsilon = 0.001$).

Figure 4: **Attacker's $\epsilon$-MSNE Strategy vs. Node Degree:** Average indegree (top) and outdegree (bottom) of nodes potentially attacked in terms of the $\epsilon$-MSNE.

**Relation to network structure.** Figure 4 presents some experimental results on the relationship between network structure and the attacker's equilibrium behavior. The graphs show, for each $\epsilon$ value, the average indegree and outdegree, across the ten BRGD runs, of those nodes that are potential targets of a direct attack at an $\epsilon$-MSNE. In general, both the average indegree and outdegree of the nodes that are potential targets of a direct attack tend to increase as $\epsilon$ decreases. One possible reason for this finding could be the fact that the values of $\alpha_i$ generated for each player are relatively low (i.e., uniformly distributed over $\left[0, \frac{1}{40}\right]$); yet, interestingly, such behavior and pattern, is exact opposite of the theory for the case $\alpha_i = 1$.